\documentclass[aps,pra,twocolumn,superscriptaddress]{revtex4-1}
\usepackage{amsmath,amsfonts,amssymb}
\usepackage{graphicx,subfigure}
\usepackage{bbm}

\begin{document}

\title{Phase-dependent which-way information}

\author{Uwe Schilling}
\affiliation{Institut f\"{u}r Optik, Information und Photonik and Erlangen Graduate School in Advanced Optical Technologies (SAOT), Universit\"{a}t Erlangen-N\"{u}rnberg, 91058 Erlangen, Germany}

\author{Joachim von Zanthier}
\affiliation{Institut f\"{u}r Optik, Information und Photonik and Erlangen Graduate School in Advanced Optical Technologies (SAOT), Universit\"{a}t Erlangen-N\"{u}rnberg, 91058 Erlangen, Germany}

\date{\today}

\begin{abstract}
We introduce a new observable for reading out a which-way detector in a Young-type interferometer whose eigenstates either contain full which-way information or none at all. We calculate the which-way knowledge $\mathcal K$ that can be retrieved from this observable and find that $\mathcal K$ depends on the phase difference $\delta$ that the interfering object accumulates on its way from either slit to the detector. In particular, it turns out that $\mathcal K(\delta)$ has an upper bound of 1, almost independent of the visibility $\mathcal V$ of the interference pattern generated by the interfering object on a screen, which is in marked contrast to the well-known inequality $\mathcal K^2 + \mathcal V^2 \leq 1$ (cf. B.-G. Englert, Phys. Rev. Lett. \textbf{77}, 2154 (1996)).
\end{abstract}


\maketitle

\section{Introduction}
Which-way detection schemes \cite{Wootters:1979,Scully:1991,Mandel:1991,Englert:1996,Duerr:1998,Buks:1998,Bjoerk:1998,Pryde:2004,Peng:2005,Kolar:2007,Jacques:2008,Barbieri:2009,Erez:2009} in a two-way interferometer and quantum erasure \cite{Scully:1982,Hillery:1983,Scully:1991,Kwiat:1992,Herzog:1995,Kim:2000,Englert:2000} are two strongly related phenomena. In both cases the setup consists of an interferometer featuring an auxiliary quantum system which detects the path of the interfering quantum mechanical object (quanton). If the experimenter strives to measure an observable whose eigenvalues can be correlated to the path of the quanton, he implements a which-way detection scheme while in case he chooses to measure eigenvalues of an observable which always projects the which-way detector (WWD) into a state that renders the acquisition of which-way (WW) information impossible a quantum eraser has been realized. In this paper, we present a scheme which can be considered a ``mixture'' of a WW detection scheme and a quantum eraser. For this purpose we introduce a new observable to be read out from the WWD whose eigenstates either carry full WW information or none at all. We show that this observable has the interesting property that the WW information it carries is correlated with the position at which the quanton is detected on the screen. In particular, with the quanton being detected in a certain region around the minima of the interference pattern the new observable provides \emph{more} WW information than the observable recognized in \cite{Englert:1996} to be optimal, while in the maxima it carries less.

The paper is organized as follows: Sec. \ref{sec_quantitative_wpd} reiterates the well-known method of quantifying wave-particle duality which will be used throughout the paper. In Sec. \ref{sec_double_slit}, a double slit interferometer with a simple WWD made up of two qubits is investigated and the new observable is introduced. Sec. \ref{sec_micromaser} discusses an exemplified application in a variation of the well-known micromaser setup by Scully \emph{et al.} \cite{Scully:1991} and we conclude in Sec. \ref{sec_conclusion}.

\section{Basic concepts of quantitative wave-particle duality}
\label{sec_quantitative_wpd}
\begin{figure}
\centering
\includegraphics[scale=1.0]{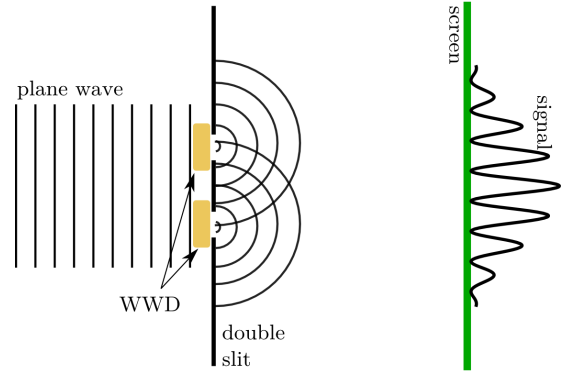}
\caption{Simplified scheme of a young-type double-slit interferometer, in which a which-way detector (WWD) is inserted in order to measure the path of the quanton.}
\label{fig_young}
\end{figure}

In a general symmetric Young-type interferometer, the wavefunction of a quanton splits at two narrow slits into the coherent parts $|\psi_a\rangle$ and $|\psi_b\rangle$ which describe the quanton emerging from the respective slit and propagating towards a screen where the quanton is registered. On the propagation, a relative phase $\delta:=\delta(\vec{r})$ is accumulated between $|\psi_a\rangle$ and $|\psi_b\rangle$ which depends on the position $\vec{r}$ at the screen where the quanton is registered. The probability of detecting a quanton at a position $\vec{r}$ corresponding to a certain value of $\delta$ can be calculated as the overlap of the initial wavefunction $|\psi\rangle = \frac{1}{\sqrt{2}}(|\psi_a\rangle + |\psi_b\rangle)$ with its phase shifted version $|\psi_\delta\rangle = \frac{1}{\sqrt{2}}(|\psi_a\rangle + e^{i\delta}|\psi_b\rangle)$:
\begin{align}
P(\delta) = |\langle \psi_\delta|\psi\rangle|^2 = \tfrac{1}{2}(1+\cos\delta).
\end{align}
As already proposed by Einstein, one may now introduce a WWD into the path of the quanton in order to determine the path it took. In the optimal case, this WWD performs a quantum non-demolition measurement in which, without any backaction on the quanton, the WWD is transferred from its initial state $|\chi^{(i)}\rangle$ into (in general) non-orthogonal states $|\chi_a\rangle$ and $|\chi_b\rangle$ which mark the path of the quanton:
\begin{multline}
(|\psi_a\rangle + |\psi_b\rangle) \otimes |\chi^{(i)}\rangle \stackrel{\text{WWD}}{\rightarrow}\\
(|\psi_a\rangle |\chi_a\rangle + |\psi_b\rangle |\chi_b\rangle ) = |\Psi^{(f)}\rangle.
\label{eq_WW_after_PS}
\end{multline}
The state of the quanton after it has passed the WWD is now described by the density matrix $\rho_Q =\text{tr}_{WWD}(|\Psi^{(f)}\rangle \langle \Psi^{(f)}|)$ obtained by tracing over the degrees of freedom of the WWD. Accordingly, the probability of detecting the quanton under a certain phase shift $\delta$ is given by the overlap of $\rho_Q$ with $|\psi_\delta\rangle$
\begin{align}
P(\delta) = \langle \psi_\delta| \rho_Q|\psi_\delta\rangle
= \frac{1}{2}(1+\mathcal V \cos (\delta + \theta)),
\label{eq_probability}
\end{align}
resulting in an interference pattern on the screen. The visibility $\mathcal V$ of this interference pattern is hereby given by
\begin{align}
\mathcal V = \frac{P_{a,\text{max}}-P_{a,\text{min}}}{P_{a,\text{max}}+P_{a,\text{min}}}= |\langle \chi_a | \chi_b\rangle |,
\label{eq_visibility}
\end{align}
and the constant phase shift $\theta$ is determined by the phase of $\langle\chi_a|\chi_b\rangle$. The result for the visiblity is not unexpected, since the smaller the overlap between $|\chi_a\rangle$ and $|\chi_b\rangle$, the easier it is to distinguish between these two quantum states and thus between the paths of the quanton. For example, if $|\chi_a\rangle$ and $|\chi_b\rangle$ are orthogonal, then reading out an observable of the WWD with these eigenvectors will give full WW information. However, one might as well (among a myriad of other possibilities) choose to read out an observable with the eigenvectors $\frac{1}{\sqrt{2}}(|\chi_a\rangle \pm |\chi_b\rangle)$. These vectors contain no WW information at all and can serve to realize a quantum eraser \cite{Scully:1991}. Furthermore, in general, $|\chi_a\rangle$ and $|\chi_b\rangle$ need not be orthogonal and consequently do not represent two different eigenstates of any observable. Thus, in order to encompass all cases, we assume that a general observable $\hat{W}$ with the eigenbasis $W = \{|W_k\rangle\}$ ($k=1,\ldots,n$) is read out from the WWD. The subsequent readout process of $\hat{W}$ will then provide information about the path of the quanton which will be, in most cases, incomplete. If one uses this information from the WWD to guess the path of the quanton, then the likelihood $\mathcal L$ of guessing the way correctly constitutes a measure for the completeness of the WW information~\cite{Englert:1996}: if no WW information is available, the photon may have taken either path and any guess is a random choice which will be right with a likelihood of $\mathcal L = 0.5$, while for full WW information, we know the way for sure and thus can pick the correct way with a likelihood of 1.

In order to calculate $\mathcal L$, we note that it is determined by two probabilities: first we need to read out the observable $\hat{W}$ of the WWD and have a certain probability $p_i$ of finding it in any of its eigenstates $|W_i\rangle$. The second probability describes the probability $q_i$ of guessing the way correctly with the detector being found in a certain state. Taking these factors into account, the likelihood of guessing the way correctly in an arbitrary run of the experiment is given by \cite{Englert:1996}:
\begin{align}
\mathcal L = \sum_i p_i q_i.
\label{eq_likelihood}
\end{align}
The probability $p_i$ of finding a certain eigenstate is determined by
\begin{align}
p_i = \langle W_i | \rho_D | W_i \rangle,
\label{eq_p_i}
\end{align}
where the density matrix $\rho_D$ denotes the state of the WWD after passage of the quanton. Assuming that such a measurement is realized and the WWD is projected into a certain eigenstate $|W_i\rangle$, one can look at the relative contributions of the two possible pathways to this final state and choose the one which contributes more~\cite{Englert:1996}. Thus, we can express the probability of guessing the way correctly from $|W_i\rangle$ as:
\begin{align}
q_i = \frac{\max\left( \left|\langle\chi_a | W_i \rangle\right|^2,\left|\langle \chi_b | W_i \rangle\right|^2 \right)}{\left|\langle \chi_a | W_i \rangle\right|^2 + \left|\langle \chi_b | W_i \rangle\right|^2}.
\label{eq_q_i}
\end{align}
$\mathcal L$ is a measure for the amount of WW information that is \emph{extracted} from the WWD by measuring $\hat{W}$: If $\mathcal L = 0.5$ ($\mathcal L = 1$), we have to make a random guess about the path of the quanton (know the way for sure). In order to have a measure which has a value of 0 (1) for no (full) WW information, one can introduce the knowledge $\mathcal K$ about the path~\cite{Englert:2000} which we will write as $\mathcal K_W$ to highlight its dependence on the eigenbasis $W$ of $\hat{W}$. $\mathcal K_W$ rescales $\mathcal L$ to the interval from 0 to 1:
\begin{align}
\mathcal K_W = 2 \mathcal L - 1 =2 \sum_i p_i q_i - 1.
\label{eq_knowledge}
\end{align}
Maximizing $\mathcal K_W$ with respect to $W$ leads to a certain optimal knowledge $\mathcal K_{\text{opt}}$ which quantifies the amount of WW information that is principally \emph{available}. The central problem of this approach is the optimization involved in finding $\mathcal K_{\text{opt}}$ and it was solved in \cite{Englert:1996}. There, it was shown that $\mathcal K_W$ is maximized if an observable $\hat{E}$ of the WWD is read out whose eigenbasis constitutes what we will call the canonical basis. It consists of the eigenvectors of the operator
\begin{align}
\hat{O} = ||\chi_a\rangle \langle \chi_a|-|\chi_b\rangle\langle \chi_b||.
\label{eq_opt_basis}
\end{align}
In addition, it was shown that the duality relation
\begin{align}
\mathcal K_W^2 + \mathcal V^2 \leq 1
\label{eq_duality}
\end{align}
always holds and is fulfiled optimally, if the WWD is initially in a pure state and the canonical observable $\hat{E}$ is read out.

\section{A Doubleslit with a simple WWD}
\label{sec_double_slit}
As an example for this procedure, let us start the investigation with a WWD which is made up of two qubits with the basis states $|0\rangle_{a,b}$ and $|1\rangle_{a,b}$, one in each path of the quanton. These qubits are initialized in the state $|0\rangle_{a,b}$. On interaction with the quanton, their state is transfered into the coherent superposition $\alpha |0\rangle_{a,b} + \beta |1\rangle_{a,b}$, so that $|\chi_a\rangle$ and $|\chi_b\rangle$ (which describe the state of the WWD if the quanton has taken the respective path) are given by
\begin{align}
\begin{array}{l}
|\chi_a\rangle = \alpha |00\rangle + \beta |10\rangle \text{\,\,\,\, and}\\
|\chi_b\rangle = \alpha |00\rangle + \beta |01\rangle
\end{array}
\end{align}
where $\beta$ governs the strength of interaction between the WWD and the quanton and the first (second) entry in the kets represents the state of the qubit in path $a$ ($b$). The state of the quanton-WWD system after the interaction and after passage of the quanton through both slits can then be described by
\begin{align}
|\Psi\rangle =\tfrac{1}{\sqrt{2}}\left( (\alpha |00\rangle + \beta |10\rangle)|\psi_a\rangle  + (\alpha |00\rangle + \beta |01\rangle)|\psi_b\rangle \right)
\label{eq_system}
\end{align}
According to Eq.~(\ref{eq_visibility}), the visibility $\mathcal V$ of the interference pattern is given by the overlap of the two possible final detector states
\begin{align}
\mathcal V = |\langle\chi_a|\chi_b\rangle| = |\alpha|^2.
\end{align}

We can now evaluate the WW knowledge one gains by reading out the WWD. In a first attempt, we opt for a natural choice and read out each qubit locally, i.e. we read out the observable $\hat{N}$ with the eigenbasis $N=\{|00\rangle,|10\rangle,|01\rangle\}$ from the WWD; we will call this the natural basis. Obviously, all WW knowledge is erased if we find the detector in state $|00\rangle$, while full WW knowledge becomes available if we find either one of the other basis states. Therefore, one may consider reading out this observable a ``mixture'' between a quantum eraser scheme and the optimal which-way detection scheme. This fixes the values of the $q_i$ to
\begin{align}
q_{00} = 0.5 \hspace{2em} q_{01} = q_{10} = 1,
\end{align}
while the probability to find the WWD in either state if the state of the whole system is described by Eq.~(\ref{eq_system}) equals
\begin{align}
p_{00} = |\alpha|^2 \hspace{2em} p_{01} = p_{10} = \frac{|\beta|^2}{2}.
\end{align}
According to Eq.~(\ref{eq_knowledge}), $\mathcal K_N$ is then given by
\begin{align}
\mathcal K_N = |\beta|^2,
\end{align}
which is less then $\sqrt{1-\mathcal V^2}$, the limiting value which could be reached according to Ref. \cite{Englert:1996}.

Next, we turn to the observable which is known to be optimal: diagonalizing the operator $\hat{O}$ from Eq.~(\ref{eq_opt_basis}) for the present WWD leads to the canonical basis with the following three vectors
\begin{align}
\begin{array}{rl}
|E_0\rangle &= \sqrt{\frac{1-|\alpha|^2}{1+|\alpha|^2}}\left( |00\rangle - \tfrac{\alpha^*}{\beta^*} \left(|10\rangle + |01\rangle  \right)\right),\\
|E_a\rangle &= \sqrt{\frac{|\alpha|^2}{1+|\alpha|^2}}\left( |00\rangle + \omega_+ |10\rangle + \omega_- |01\rangle \right),\\
|E_b\rangle &= \sqrt{\frac{|\alpha|^2}{1+|\alpha|^2}}\left( |00\rangle + \omega_- |10\rangle + \omega_+ |01\rangle \right),
\end{array}
\label{eq_englerts_basis}
\end{align}
with $\omega_\pm = \tfrac{|\beta|^2 \pm \gamma}{2\alpha\beta^*}$ and $\gamma = \sqrt{1-|\alpha|^4}$. The probability of finding a certain state is given by (cf. Eq.~(\ref{eq_p_i}))
\begin{align}
p_{E_0}= 0\,\, \text{ and }\,\,
p_{E_a}=p_{E_b}=\frac{1}{2},
\end{align}
while the probability $q_i$ of guessing the way correctly from that state is given by (cf. Eq.~(\ref{eq_q_i}))
\begin{align}
q_{E_0}=\frac{1}{2}\,\, \text{ and }\,\,
q_{E_a}=q_{E_b}=\frac{1}{2}(1+\gamma).
\end{align}
Using Eq.~\ref{eq_knowledge}, the knowledge about the path is then calculated to be
\begin{align}
\mathcal K_E = \sqrt{1-|\alpha|^4} = \sqrt{1-\mathcal V^2}.
\end{align}
As expected, $\mathcal K_E$ fulfills Eq.~(\ref{eq_duality}) optimally, i.e. $\mathcal K_E^2 + \mathcal V^2 = 1$. Thus, in general, if one aims to maximize the WW information, it would be preferable to read out $\hat{E}$ over $\hat{N}$.

However, the approach as laid out so far takes into account the correlations between the quanton and the WWD to make statements only about which slit the quanton has passed through. Not accounted for is that the entangled state in Eq.~(\ref{eq_WW_after_PS}) can also lead to correlations between the detection probabilities of eigenstates of an observable of the WWD and the detection positions of the quanton on the screen. To elucidate this point, let us investigate how much we can know about the path of quantons being diffracted towards a certain position on the screen, i.e., for quantons having acquired a certain phase shift $\delta$. The joint probability $p_i(\delta)$ of finding a quanton with a phase shift of $\delta$ and the WWD in its eigenstate $|W_i\rangle$ is calculated as
\begin{align}
p_i(\delta) = |\langle W_i|\langle \psi_\delta |\Psi^{(f)}\rangle|^2.
\label{eq_joint_prob}
\end{align}
Consequently, for a quanton being detected at a position on the screen corresponding to a phase shift of $\delta$, $\mathcal K_W(\delta)$ is given by:
\begin{align}
\mathcal K_W(\delta) = 2\frac{\sum_i q_i p_i(\delta)}{\sum_i p_i(\delta)}-1.
\label{eq_L_of_delta}
\end{align}
Naturally, $\mathcal K_W(\delta)$ is gouverned on the one hand by the WW information $q_i$, contained in the eigenstates $|W_i\rangle$ of the observable $\hat{W}$, and on the other hand by the joint probability $p_i(\delta)$. For the canonical observable $\hat{E}$, one finds that $\mathcal K_E(\delta)$ is given by
\begin{align}
\mathcal K_E(\delta) = \sqrt{1-\mathcal V^2} = \mathcal K_E,
\end{align}
i.e., the path knowledge is not correlated with $\delta$, so that the same WW information is available for quantons at any point on the screen. However, for the natural observable, we arrive at a different result. For the observable $\hat{N}$, the knowledge about the path of the quantons having acquired a phase shift of $\delta$ is given by
\begin{align}
\mathcal K_N(\delta) = \frac{1-\mathcal V}{1+\mathcal V \cos \delta}.
\end{align}
\begin{figure}
\centering
\includegraphics[scale=0.17]{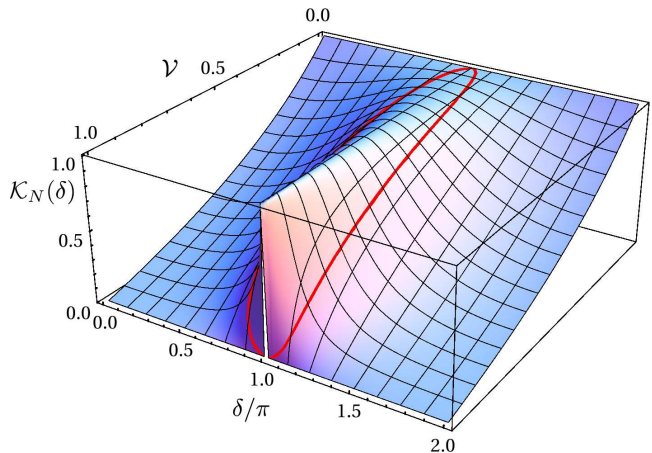}
\caption{$\mathcal K_N(\delta)$ plotted as a function of the phase shift $\delta$ (in units of $\pi$) and the visibility $\mathcal V$. The thick red line marks the region in which the natural observable allows to gain more WW information from the WWD than the canonical observable.}
\label{fig_final}
\end{figure}
This result is plotted in Fig.~\ref{fig_final}. It is interesting to note that the knowledge about the path is now indeed correlated with the position on the screen where the quanton is detected. In particular, for all quantons arriving in the minima of the interference pattern $P(\delta)$, i.e., at $\cos \delta = -1$, one always has full WW knowledge, regardless of the value of the visibility \footnote{In the particular case $\mathcal V = 1$, there is of course a vanishing probability to detect a quanton in the minima of the interference pattern.} (cf. Fig.~\ref{fig_final}). Moreover, while the observable $\hat{E}$ performs better on average, for all quantons arriving in a region around those minima, the observable $\hat{N}$ provides more WW information than $\hat{E}$. To be more exact, this is true for quantons for which the phase shift $\delta$ fulfills the following relation:
\begin{align}
\cos \delta < \frac{\sqrt{\frac{1-\mathcal V}{1+\mathcal V}}-1}{\mathcal V}.
\end{align}

This can be understood in two ways. One way is to look at what happens to the quanton wavefunction when the WWD is read out. The other way is to look at the WWD wavefunction if the quanton is detected.

In the first case, one can clearly see from Eq.~(\ref{eq_system}) that the detection of the WWD in either state $|10\rangle$ or $|01\rangle$ projects the quanton into state $|\psi_a\rangle$ or $|\psi_b\rangle$, respectively. Thus, the quanton is emerging exactly from one slit, in which case no interference occurs and one has a constant probability of detecting the quanton anywhere on the screen. However, if the WWD is found to be in state $|00\rangle$, the wavefunction of the quanton is projected into the state $\frac{1}{\sqrt{2}}(|\psi_a\rangle + |\psi_b\rangle)$, i.e. an equal superposition of both paths, in which case all WW information has been erased. In this case, i.e., with the detector being found in $|00\rangle$, no quantons arrive in the minima of the interference pattern and only very few in their vicinity due to destructive interference between $|\psi_a\rangle$ and $|\psi_b\rangle$ around $\delta=\pi$. Therefore, all of the quantons that actually are detected in the minima of an interference pattern with $\mathcal V <1$ and most of the ones in the surroundings coincide with a WWD state which carries full WW information.

The second way of understanding the correlations is by looking at the WWD wavefunction upon detection of a quanton at the screen. After the detection of a quanton with phase shift $\delta$, one finds the state of the WWD to be projected into the state
\begin{multline}
|\chi_\delta\rangle = \frac{1}{\zeta(\delta)}\left(|\chi_a\rangle + e^{i\delta}|\chi_b\rangle\right)\\
= \frac{1}{\zeta(\delta)}\left(\alpha(1+e^{i\delta})|00\rangle + \beta |10\rangle + \beta e^{i\delta}|01\rangle\right)
\label{eq_projected_detector_state}
\end{multline}
with $\zeta(\delta)$ determined by the normalization. From Eq.~(\ref{eq_projected_detector_state}), we can see that whenever a quanton is detected in the vicinity of $\delta=\pi$, there is only a tiny chance of finding the WWD in the state $|00\rangle$, whereas one is almost certain to find the WWD either in state $|01\rangle$ or $|10\rangle$. This is a result of an interference effect in the WWD which cancels the overlapping parts of $|\chi_a\rangle$ and $|\chi_b\rangle$, $\alpha|0\rangle$, so that only the non-overlapping parts, $\beta |10\rangle$ and $\beta |01\rangle$, contribute to the final WWD wavefunction, both determining unambigouosly the path of the quanton.

For the present case, we have chosen a very simple type of WWD, namely two qubits. This was done for didactic reasons only. In the appendix, we show that a natural observable with exactly the properties described above always exists regardless of the specific implementation of the WWD and give explicit expressions for the natural basis in terms of $|\chi_{a,b}\rangle$.

\section{Example: the micromaser}
\label{sec_micromaser}
\begin{figure}
\centering
\includegraphics[scale=0.48]{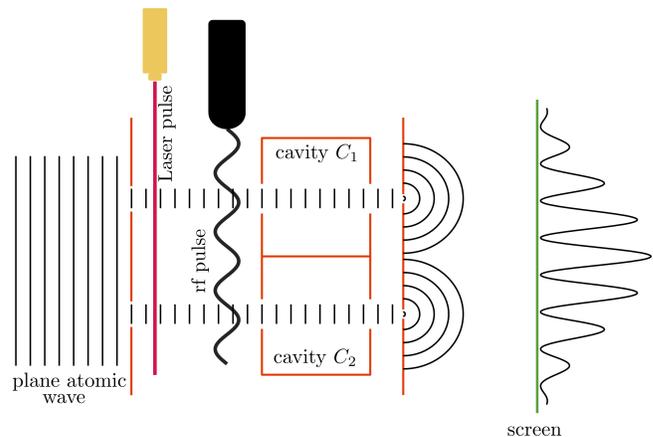}
\caption{Simplified scheme of the micromaser setup: atoms enter as a plane wave from the left. Two wide appertures seperate the atomic beam spacially into two collimated beams. Then, a laser excites the atoms into a long-lived Rydberg state $|x\rangle$ and an rf-pulse transfers them from there into a superposition of two Rydberg states $\cos\theta |x\rangle + \sin \theta |y\rangle$. The cavities are tuned to the resonance $|x\rangle \rightarrow |y\rangle$ and their parameters are adjusted such that $|y\rangle$ deexcites with certainty under emission of a photon. After passing the cavities, the atoms are diffracted at two narrow slits and their interference pattern is observed on a screen.}
\label{fig_micromaser}
\end{figure}

The effects described in the previous chapter can be found in all Young-type interferometers. However, as a specific example, let us look at a variation of the famous quantum eraser setup by Scully \emph{et al.} \cite{Scully:1991}, shown in Fig. \ref{fig_micromaser}. Disregarding the laser-pulse, the rf-pulse and the cavities from Fig. \ref{fig_micromaser} for the moment, the setup forms a basic atom interferometer: A plane atomic wave hits a set of wider slits where the beam is collimated and then illuminates two narrow slits where the wave is diffracted. The wave function describing the center of mass motion of the atom after the second double slit is then given by a sum of the two wavefunctions originating at the two narrow slits:
\begin{align}
\Psi(\vec{r}) = \frac{1}{\sqrt{2}}[\psi_1(\vec{r}) + \psi_2(\vec{r})].
\end{align}
The probability to detect an atom at a certain position $\vec{r}$ on the screen is then given by:
\begin{multline}
P(\vec{r}) = \frac{1}{2}\left[|\psi_1(\vec{r})|^2 + |\psi_2(\vec{r})|^2\right.\\
\left.+ \psi_1^*(\vec{r})\psi_2(\vec{r}) + \psi_2^*(\vec{r})\psi_1(\vec{r})\right].
\end{multline}
The interference is given as usual by the cross terms $\psi_1^*\psi_2 + \psi_2^*\psi_1$. Now we include a laser which excites a long lived Rydberg state $|x\rangle$ in the atoms and an rf-pulse which transfers the internal state of the atom into the coherent superposition $\cos \theta |x\rangle + \sin \theta |y\rangle$ of $|x\rangle$ and an energetically higher lying Rydberg state $|y\rangle$. Behind the laser and the rf-pulse generator, we place two identical cavities $C_1$ and $C_2$, arranged as shown in Fig. \ref{fig_micromaser}. The cavities -- both initially in the vacuum state $|0\rangle$ -- are assumed to be tuned into resonance with the transition of the internal atomic state $|x\rangle \rightarrow |y\rangle$ and their parameters are trimmed such that exactly half a Rabi oscillation takes place for the atomic state $|y\rangle$ when an atom is passing through the cavities. As a result, after the passage of the atom through $C_1$ and $C_2$, the internal atomic state is $|x\rangle$, while the cavities are left in a superposition of either containing a photon or not, thus marking the path of the atom:
\begin{multline}
\Psi(\vec{r}) = \frac{1}{\sqrt{2}}\left[\psi_1(\vec{r})(\cos \theta |0 0\rangle + \sin \theta |1 0\rangle)\right.\\
\left.+ \psi_2(\vec{r})(\cos \theta |0 0\rangle + \sin \theta |0 1\rangle)\right]|y\rangle,
\end{multline}
where the kets $|C_1 C_2\rangle$ denote the number of photons in the cavities $C_1$ and $C_2$, respectively. The probability to find an atom at point $\vec{r}$ is now given by
\begin{multline}
P(\vec{r}) = \frac{1}{2}\left[|\psi_1(\vec{r})|^2 + |\psi_2(\vec{r})|^2\right.\\
\left.+ \cos^2 \theta \psi_1^*(\vec{r})\psi_2(\vec{r}) + \cos^2 \theta \psi_2^*(\vec{r})\psi_1(\vec{r})\right],
\end{multline}
where the visibility $\mathcal V$ of the interference pattern is reduced to $\mathcal V = \cos^2 \theta$. At the same time, the detection of an atom at position $\vec{r}$ projects the cavities into the state
\begin{align}
|C_1 C_2\rangle = \cos \theta (1+e^{i\delta}) |0 0\rangle
+ \sin \theta |1 0\rangle + \sin \theta e^{i\delta} |0 1\rangle.
\label{eq_maser_final_state}
\end{align}
In this case, the natural basis is simply made up of the Fock number states $|00\rangle$, $|01\rangle$, and $|10\rangle$ of the cavities, i.e. one simply checks whether one finds a photon in either cavity. If one does find a photon, one knows that the atom has passed through the respective cavity, while if one finds both cavities to be empty, all WW information contained in the state of the WWD described by Eq.~(\ref{eq_maser_final_state}) has been erased. As discussed in Sec. \ref{sec_double_slit}, the natural basis provides more WW information than the canonical basis for all atoms arriving close to the minima of the interference pattern, while the canonical basis performs better over a wider range around the maxima and also on average over a large number of runs of the experiment.

\section{Conclusion}
\label{sec_conclusion}
In conclusion, we have shown that while it is not possible to beat the canonical observable in terms of average WW information, the natural observable does lead to more WW information for a certain subset of quantons passing through a Young-type interferometer. In addition and especially for high values of the visibility, there is an automatic sorting taking place: the less which-way information the WWD reveals, the more likely it is to find the quanton close to the maxima of the interference pattern; thus, we have realized what might be called a `self-sorting quantum eraser'. In addition, we were able to show by explicit construction that a natural observable exists for all WWD. Further work might include the extension of the analysis to asymmetric interferometers and mixed WWD states.
The authors gratefully acknowledge funding of the Erlangen Graduate School in Advanced Optical Technologies (SAOT) by the German Research Foundation (DFG) in the framework of the German excellence initiative. US thanks the Elite Network of Bavaria for financial support.

\appendix
\section{Introducing a general WWD}
\label{sec_general_wwd}
In Sec. \ref{sec_double_slit}, we have seen that for a certain type of WWD (two identical qubits), one finds an observable of the WWD, which allows to identify the path of the particles arriving in the minima of the interference pattern with certainty, independent of the visibility of the interference pattern. We will now tackle the question whether such an observable, i.e. a natural basis, exists for all types of WWDs.

First we note that in order to find a basis in which one state can be identified uniquely with either of two paths while a third state carries no WW information, the Hilbert space $\mathcal H$ of the WWD needs to have at least three dimensions. Thus, if the WWD is only a qubit, one cannot find a natural basis. However, in such a case, one may add an auxiliary qubit which does not take part in the detection process but which is also read out. In this way, $\mathcal H$ is enlarged to four dimensions. Therefore, we can assume that dim $\mathcal H$ $\geq 3$ always holds. In such a Hilbert space, the vectors $|\chi_a\rangle$ and $|\chi_b\rangle$ span a two-dimensional subspace (excluding the limiting case where $|\langle \chi_a|\chi_b\rangle|=1$), and there exists thus a vector $|t\rangle \in \mathcal H$ which is orthogonal to both $|\chi_a\rangle$ and $|\chi_b\rangle$. In the following, we will construct an orthonormal natural basis $\{|0\rangle,|+\rangle,|-\rangle\}$ which spans the same subspace as the vectors $\{|\chi_a\rangle,|\chi_b\rangle,|t\rangle\}$.
\begin{figure}
\centering
\includegraphics[scale=0.6]{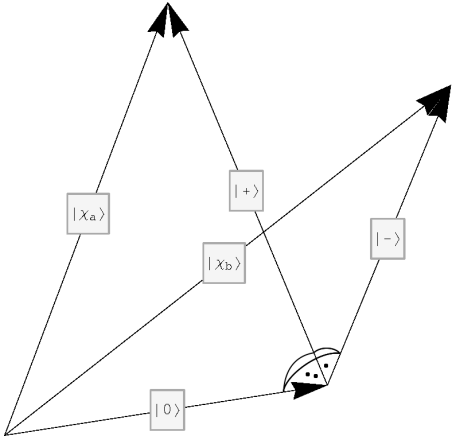}
\caption{The figure shows what a natural basis would look like in real vector space. It visualizes the approach for the construction of the natural basis. The states $|0\rangle$, $|-\rangle$, and $|+\rangle$ are pairwise orthogonal.}
\label{fig_vectors}
\end{figure}

In a first step, $|0\rangle$ is expressed as a sum of the three known vectors:
\begin{align}
|0\rangle = x_1 |\chi_a\rangle + x_1 e^{-i \gamma} |\chi_b\rangle + x_2 |t\rangle,
\label{eq_0}
\end{align}
with $x_1$, $x_2$, and $\gamma$ to be determined. Since this vector is supposed to carry no WW information, the prefactors of $|\chi_a\rangle$ and $|\chi_b\rangle$ can at most differ by a relative phase $\gamma$ (cf. Eq.~(\ref{eq_q_i})). In addition, we want $|\chi_a\rangle$ ($|\chi_b\rangle$) to be orthogonal to $|-\rangle$ ($|+\rangle$), because the states $|\pm\rangle$ are supposed to carry full WW information, i.e., overlap only with either $|\chi_a\rangle$ or $|\chi_b\rangle$. Thus, we can write them as
\begin{align}
\begin{array}{l}
|\chi_a\rangle = v|0\rangle + \sqrt{1-|v|^2} |+\rangle\\
|\chi_b\rangle = v e^{i\gamma}|0\rangle + \sqrt{1-|v|^2} |-\rangle.
\end{array}
\label{eq_chis}
\end{align}
Fig.~\ref{fig_vectors} shows how these conditions can be fulfiled in a real threedimensional vector space. What is done in the following, is essentially to construct the corresponding vectors in the complex Hilbert space.

From Eq.~(\ref{eq_0}), it follows immediately that the weight of $|0\rangle$ in $|\chi_a\rangle$ and $|\chi_b\rangle$ has to have the same absolute value, while the phase $\gamma$ and the absolute value of $v$ are directly defined by the overlap of $|\chi_a\rangle$ and $|\chi_b\rangle$ (remember that $|\chi_{a,b}\rangle$ are given)
\begin{align}
\langle \chi_a | \chi_b \rangle = |v|^2 e^{i\gamma}.
\label{eq_v}
\end{align}
Pairwise orthogonality of $|0\rangle$, $|+\rangle$, and $|-\rangle$ and the decomposition of $|\chi_a\rangle$ and $|\chi_b\rangle$ introduced in Eq.~(\ref{eq_chis}) lead to the following relations:
\begin{subequations}
\begin{align}
|\chi_a\rangle - v |0\rangle &\perp |0\rangle\\
|\chi_b\rangle - v e^{i\gamma}|0\rangle &\perp |0\rangle\\
|\chi_a\rangle - v |0\rangle &\perp |\chi_b\rangle - v e^{i\gamma}|0\rangle.
\end{align}
\end{subequations}
These conditions can be rewritten in the form of scalar products:
\begin{subequations}
\begin{align}
\langle \chi_a|0\rangle - v^* \langle 0|0\rangle = 0
\label{eq_po0}\\
\langle \chi_b|0\rangle - v^* e^{-i\gamma} \langle 0|0\rangle = 0
\label{eq_mo0}\\
\langle \chi_a|\chi_b\rangle -v^* \langle 0|\chi_b\rangle - v e^{i\gamma}\underbrace{\left(\langle \chi_a | 0 \rangle - v^* \langle 0|0\rangle\right)}_{=0\text{ cf. Eq. (\ref{eq_po0})}} = 0
\end{align}
\end{subequations}
Using the notation introduced in Eq.~(\ref{eq_0}) and Eq.~(\ref{eq_v}) the scalar products can be expanded:
\begin{subequations}
\begin{align}
x_1 + |v|^2 x_1 - v^* = 0,
\label{eq_po02}\\
|v|^2 x_1 + x_1 - v^* = 0,
\label{eq_mo02}\\
|v|^2 -v^*(1 + |v|^2) x_1^* = 0,
\label{eq_pom2}\\
|x_1|^2 + |x_2|^2 = 1,
\label{eq_norm}
\end{align}
\end{subequations}
where the first three equations are obviously pairwise equivalent, while the last equation simply denotes the normalization of $|0\rangle$. Since $x_2$ appears only as the square of an absolute value, it will be determined only up to a phase.
The value of $x_1$ can be deduced from any of the first three equations above by solving for $x_1$:
\begin{align}
x_1 = \frac{v^*}{1 + |v|^2}.
\end{align}
Then, Eq.~(\ref{eq_norm}) can be used to derive $x_2$:
\begin{align}
|x_2|^2 = \frac{1 + |v|^4}{(1 + |v|^2)^2}
\hspace{2em}\Rightarrow \hspace{2em}x_2 = e^{i\eta}\frac{\sqrt{1 + |v|^4}}{1+|v|^2}
\end{align}
with $\eta$ arbitrary. Thus, for given $|\chi_a\rangle$ and $|\chi_b\rangle$, the vectors of the natural basis have the following form:
\begin{subequations}
\begin{align}
|0\rangle = \frac{v^*|\chi_a\rangle + e^{-i\gamma}v^*|\chi_b\rangle + e^{i\eta}\sqrt{1 + |v|^4} |t\rangle}{1+|v|^2}\\
|+\rangle = \frac{|\chi_a\rangle - |v|^2 e^{-i\gamma}|\chi_b\rangle - v e^{i\eta}\sqrt{1 + |v|^4} |t\rangle}{(1+|v|^2)\sqrt{1-|v|^2}}\\
|-\rangle = \frac{|\chi_b\rangle - |v|^2 e^{i\gamma}|\chi_a\rangle - v e^{i\gamma} e^{i\eta}\sqrt{1 + |v|^4} |t\rangle}{(1+|v|^2)\sqrt{1-|v|^2}}
\end{align}
\end{subequations}
with $\eta$ and the phase of $v$ arbitrary. These three vectors are by construction pairwise orthogonal and at the same time $|+\rangle$ ($|-\rangle$) has no overlap with $|\chi_b\rangle$ ($|\chi_a\rangle$). The only condition necessary to construct these vectors was $|\langle\chi_a|\chi_b\rangle|\neq 1$ which simply states in mathematical terms that the WWD does collect some WW information. Therefore, a natural basis exists for any type of WWD, i.e. one can always choose an observable for reading out the WWD such that there exists full WW information for all particles arriving in the minima of the interference pattern, regardless of the structure of the WWD.

\bibliography{bibliography_long_names}

\begin{thebibliography}{10}%
\makeatletter
\providecommand \@ifxundefined [1]{%
 \ifx #1\undefined \expandafter \@firstoftwo
 \else \expandafter \@secondoftwo
\fi
}%
\providecommand \@ifnum [1]{%
 \ifnum #1\expandafter \@firstoftwo
 \else \expandafter \@secondoftwo
\fi
}%
\providecommand \enquote [1]{``#1''}%
\providecommand \bibnamefont  [1]{#1}%
\providecommand \bibfnamefont [1]{#1}%
\providecommand \citenamefont [1]{#1}%
\providecommand\href[0]{\@sanitize\@href}%
\providecommand\@href[1]{\endgroup\@@startlink{#1}\endgroup\@@href}%
\providecommand\@@href[1]{#1\@@endlink}%
\providecommand \@sanitize [0]{\begingroup\catcode`\&12\catcode`\#12\relax}%
\@ifxundefined \pdfoutput {\@firstoftwo}{%
 \@ifnum{\z@=\pdfoutput}{\@firstoftwo}{\@secondoftwo}%
}{%
 \providecommand\@@startlink[1]{\leavevmode\special{html:<a href="#1">}}%
 \providecommand\@@endlink[0]{\special{html:</a>}}%
}{%
 \providecommand\@@startlink[1]{%
  \leavevmode
  \pdfstartlink
   attr{/Border[0 0 1 ]/H/I/C[0 1 1]}%
   user{/Subtype/Link/A<</Type/Action/S/URI/URI(#1)>>}%
  \relax
 }%
 \providecommand\@@endlink[0]{\pdfendlink}%
}%
\providecommand \url  [0]{\begingroup\@sanitize \@url }%
\providecommand \@url [1]{\endgroup\@href {#1}{\urlprefix}}%
\providecommand \urlprefix [0]{URL }%
\providecommand \Eprint[0]{\href }%
\@ifxundefined \urlstyle {%
  \providecommand \doi [1]{doi:\discretionary{}{}{}#1}%
}{%
  \providecommand \doi [0]{doi:\discretionary{}{}{}\begingroup
  \urlstyle{rm}\Url }%
}%
\providecommand \doibase [0]{http://dx.doi.org/}%
\providecommand \Doi[1]{\href{\doibase#1}}%
\providecommand \bibAnnote [3]{%
  \BibitemShut{#1}%
  \begin{quotation}\noindent
    \textsc{Key:}\ #2\\\textsc{Annotation:}\ #3%
  \end{quotation}%
}%
\providecommand \bibAnnoteFile [2]{%
  \IfFileExists{#2}{\bibAnnote {#1} {#2} {\input{#2}}}{}%
}%
\providecommand \typeout [0]{\immediate \write \m@ne }%
\providecommand \selectlanguage [0]{\@gobble}%
\providecommand \bibinfo [0]{\@secondoftwo}%
\providecommand \bibfield [0]{\@secondoftwo}%
\providecommand \translation [1]{[#1]}%
\providecommand \BibitemOpen[0]{}%
\providecommand \bibitemStop [0]{}%
\providecommand \bibitemNoStop [0]{.\EOS\space}%
\providecommand \EOS [0]{\spacefactor3000\relax}%
\providecommand \BibitemShut [1]{\csname bibitem#1\endcsname}%
\bibitem{Wootters:1979}%
  \BibitemOpen
  \bibfield{author}{%
  \bibinfo {author} {\bibfnamefont{W.~K.}\ \bibnamefont{Wootters}}\ and\
  \bibinfo {author} {\bibfnamefont{W.~H.}\ \bibnamefont{Zurek}},\ }%
  \bibfield{journal}{%
  \bibinfo {journal} {Phys. Rev. D}\ }%
  \textbf{\bibinfo {volume} {19}},\ \bibinfo {pages} {473} (\bibinfo {year}
  {1979})%
  \bibAnnoteFile{NoStop}{Wootters:1979}%
\bibitem{Scully:1991}%
  \BibitemOpen
  \bibfield{author}{%
  \bibinfo {author} {\bibfnamefont{M.~O.}\ \bibnamefont{Scully}}, \bibinfo
  {author} {\bibfnamefont{B.-G.}\ \bibnamefont{Englert}},\ and\ \bibinfo
  {author} {\bibfnamefont{H.}~\bibnamefont{Walther}},\ }%
  \bibfield{journal}{%
  \bibinfo {journal} {Nature}\ }%
  \textbf{\bibinfo {volume} {351}},\ \bibinfo {pages} {111} (\bibinfo {year}
  {1991})%
  \bibAnnoteFile{NoStop}{Scully:1991}%
\bibitem{Mandel:1991}%
  \BibitemOpen
  \bibfield{author}{%
  \bibinfo {author} {\bibfnamefont{L.}~\bibnamefont{Mandel}},\ }%
  \bibfield{journal}{%
  \bibinfo {journal} {Opt. Lett.}\ }%
  \textbf{\bibinfo {volume} {16}},\ \bibinfo {pages} {1882} (\bibinfo {year}
  {1991})%
  \bibAnnoteFile{NoStop}{Mandel:1991}%
\bibitem{Englert:1996}%
  \BibitemOpen
  \bibfield{author}{%
  \bibinfo {author} {\bibfnamefont{B.-G.}\ \bibnamefont{Englert}},\ }%
  \bibfield{journal}{%
  \bibinfo {journal} {Phys. Rev. Lett.}\ }%
  \textbf{\bibinfo {volume} {77}},\ \bibinfo {pages} {2154} (\bibinfo {year}
  {1996})%
  \bibAnnoteFile{NoStop}{Englert:1996}%
\bibitem{Duerr:1998}%
  \BibitemOpen
  \bibfield{author}{%
  \bibinfo {author} {\bibfnamefont{S.}~\bibnamefont{D{\"u}rr}}, \bibinfo
  {author} {\bibfnamefont{T.}~\bibnamefont{Nonn}},\ and\ \bibinfo {author}
  {\bibfnamefont{G.}~\bibnamefont{Rempe}},\ }%
  \bibfield{journal}{%
  \bibinfo {journal} {Nature}\ }%
  \textbf{\bibinfo {volume} {395}},\ \bibinfo {pages} {33} (\bibinfo {year}
  {1998})%
  \bibAnnoteFile{NoStop}{Duerr:1998}%
\bibitem{Buks:1998}%
  \BibitemOpen
  \bibfield{author}{%
  \bibinfo {author} {\bibfnamefont{E.}~\bibnamefont{Buks}}, \bibinfo {author}
  {\bibfnamefont{R.}~\bibnamefont{Schuster}}, \bibinfo {author}
  {\bibfnamefont{M.}~\bibnamefont{Heiblum}}, \bibinfo {author}
  {\bibfnamefont{D.}~\bibnamefont{Mahalu}},\ and\ \bibinfo {author}
  {\bibfnamefont{V.}~\bibnamefont{Umansky}},\ }%
  \bibfield{journal}{%
  \bibinfo {journal} {Nature}\ }%
  \textbf{\bibinfo {volume} {391}},\ \bibinfo {pages} {871} (\bibinfo {year}
  {1998})%
  \bibAnnoteFile{NoStop}{Buks:1998}%
\bibitem{Bjoerk:1998}%
  \BibitemOpen
  \bibfield{author}{%
  \bibinfo {author} {\bibfnamefont{G.}~\bibnamefont{Bj{\"o}rk}}\ and\ \bibinfo
  {author} {\bibfnamefont{A.}~\bibnamefont{Karlsson}},\ }%
  \bibfield{journal}{%
  \bibinfo {journal} {Phys. Rev. A}\ }%
  \textbf{\bibinfo {volume} {58}},\ \bibinfo {pages} {3477} (\bibinfo {year}
  {1998})%
  \bibAnnoteFile{NoStop}{Bjoerk:1998}%
\bibitem{Pryde:2004}%
  \BibitemOpen
  \bibfield{author}{%
  \bibinfo {author} {\bibfnamefont{G.~J.}\ \bibnamefont{Pryde}}, \bibinfo
  {author} {\bibfnamefont{J.~L.}\ \bibnamefont{O'Brien}}, \bibinfo {author}
  {\bibfnamefont{A.~G.}\ \bibnamefont{White}}, \bibinfo {author}
  {\bibfnamefont{S.~D.}\ \bibnamefont{Bartlett}},\ and\ \bibinfo {author}
  {\bibfnamefont{T.~C.}\ \bibnamefont{Ralph}},\ }%
  \bibfield{journal}{%
  \bibinfo {journal} {Phys. Rev. Lett.}\ }%
  \textbf{\bibinfo {volume} {92}},\ \bibinfo {pages} {190402} (\bibinfo {year}
  {2004})%
  \bibAnnoteFile{NoStop}{Pryde:2004}%
\bibitem{Peng:2005}%
  \BibitemOpen
  \bibfield{author}{%
  \bibinfo {author} {\bibfnamefont{X.}~\bibnamefont{Peng}}, \bibinfo {author}
  {\bibfnamefont{X.}~\bibnamefont{Zhu}}, \bibinfo {author}
  {\bibfnamefont{D.}~\bibnamefont{Suter}}, \bibinfo {author}
  {\bibfnamefont{J.}~\bibnamefont{Du}}, \bibinfo {author}
  {\bibfnamefont{M.}~\bibnamefont{Liu}},\ and\ \bibinfo {author}
  {\bibfnamefont{K.}~\bibnamefont{Gao}},\ }%
  \bibfield{journal}{%
  \bibinfo {journal} {Phys. Rev. A}\ }%
  \textbf{\bibinfo {volume} {72}},\ \bibinfo {pages} {052109} (\bibinfo {year}
  {2005})%
  \bibAnnoteFile{NoStop}{Peng:2005}%
\bibitem{Kolar:2007}%
  \BibitemOpen
  \bibfield{author}{%
  \bibinfo {author} {\bibfnamefont{M.}~\bibnamefont{Kol{\'a}\v{r}}}, \bibinfo
  {author} {\bibfnamefont{T.}~\bibnamefont{Opatrn{\'y}}}, \bibinfo {author}
  {\bibfnamefont{N.}~\bibnamefont{Bar-Gill}}, \bibinfo {author}
  {\bibfnamefont{N.}~\bibnamefont{Erez}},\ and\ \bibinfo {author}
  {\bibfnamefont{G.}~\bibnamefont{Kurizki}},\ }%
  \bibfield{journal}{%
  \bibinfo {journal} {New J. Phys.}\ }%
  \textbf{\bibinfo {volume} {9}},\ \bibinfo {pages} {129} (\bibinfo {year}
  {2007})%
  \bibAnnoteFile{NoStop}{Kolar:2007}%
\bibitem{Jacques:2008}%
  \BibitemOpen
  \bibfield{author}{%
  \bibinfo {author} {\bibfnamefont{V.}~\bibnamefont{Jacques}}, \bibinfo
  {author} {\bibfnamefont{E.}~\bibnamefont{Wu}}, \bibinfo {author}
  {\bibfnamefont{F.}~\bibnamefont{Grosshans}}, \bibinfo {author}
  {\bibfnamefont{F.}~\bibnamefont{Treussart}}, \bibinfo {author}
  {\bibfnamefont{P.}~\bibnamefont{Grangier}}, \bibinfo {author}
  {\bibfnamefont{A.}~\bibnamefont{Aspect}},\ and\ \bibinfo {author}
  {\bibfnamefont{J.-F.}\ \bibnamefont{Roch}},\ }%
  \bibfield{journal}{%
  \bibinfo {journal} {Phys. Rev. Lett.}\ }%
  \textbf{\bibinfo {volume} {100}},\ \bibinfo {pages} {220402} (\bibinfo {year}
  {2008})%
  \bibAnnoteFile{NoStop}{Jacques:2008}%
\bibitem{Barbieri:2009}%
  \BibitemOpen
  \bibfield{author}{%
  \bibinfo {author} {\bibfnamefont{M.}~\bibnamefont{Barbieri}}, \bibinfo
  {author} {\bibfnamefont{M.~E.}\ \bibnamefont{Goggin}}, \bibinfo {author}
  {\bibfnamefont{M.~P.}\ \bibnamefont{Almeida}}, \bibinfo {author}
  {\bibfnamefont{B.~P.}\ \bibnamefont{Lanyon}},\ and\ \bibinfo {author}
  {\bibfnamefont{A.~G.}\ \bibnamefont{White}},\ }%
  \bibfield{journal}{%
  \bibinfo {journal} {New J. Phys.}\ }%
  \textbf{\bibinfo {volume} {11}},\ \bibinfo {pages} {093012} (\bibinfo {year}
  {2009})%
  \bibAnnoteFile{NoStop}{Barbieri:2009}%
\bibitem{Erez:2009}%
  \BibitemOpen
  \bibfield{author}{%
  \bibinfo {author} {\bibfnamefont{N.}~\bibnamefont{Erez}}, \bibinfo {author}
  {\bibfnamefont{D.}~\bibnamefont{Jacobs}},\ and\ \bibinfo {author}
  {\bibfnamefont{G.}~\bibnamefont{Kurizki}},\ }%
  \bibfield{journal}{%
  \bibinfo {journal} {J. Phys. B}\ }%
  \textbf{\bibinfo {volume} {42}},\ \bibinfo {pages} {114006} (\bibinfo {year}
  {2009})%
  \bibAnnoteFile{NoStop}{Erez:2009}%
\bibitem{Scully:1982}%
  \BibitemOpen
  \bibfield{author}{%
  \bibinfo {author} {\bibfnamefont{M.~O.}\ \bibnamefont{Scully}}\ and\ \bibinfo
  {author} {\bibfnamefont{K.}~\bibnamefont{Dr{\"u}hl}},\ }%
  \bibfield{journal}{%
  \bibinfo {journal} {Phys. Rev. A}\ }%
  \textbf{\bibinfo {volume} {25}},\ \bibinfo {pages} {2208} (\bibinfo {year}
  {1982})%
  \bibAnnoteFile{NoStop}{Scully:1982}%
\bibitem{Hillery:1983}%
  \BibitemOpen
  \bibfield{author}{%
  \bibinfo {author} {\bibfnamefont{M.}~\bibnamefont{Hillery}}\ and\ \bibinfo
  {author} {\bibfnamefont{M.~O.}\ \bibnamefont{Scully}},\ }%
  \enquote{\bibinfo {title} {Quantum optics, experimental gravity, and
  measurement theory},}\ \ (\bibinfo {publisher} {Plenum},\ \bibinfo {address}
  {New York},\ \bibinfo {year} {1983})\ p.~\bibinfo {pages} {65}%
  \bibAnnoteFile{NoStop}{Hillery:1983}%
\bibitem{Kwiat:1992}%
  \BibitemOpen
  \bibfield{author}{%
  \bibinfo {author} {\bibfnamefont{P.~G.}\ \bibnamefont{Kwiat}}, \bibinfo
  {author} {\bibfnamefont{A.~M.}\ \bibnamefont{Steinberg}},\ and\ \bibinfo
  {author} {\bibfnamefont{R.~Y.}\ \bibnamefont{Chiao}},\ }%
  \bibfield{journal}{%
  \bibinfo {journal} {Phys. Rev. A}\ }%
  \textbf{\bibinfo {volume} {45}},\ \bibinfo {pages} {7729} (\bibinfo {year}
  {1992})%
  \bibAnnoteFile{NoStop}{Kwiat:1992}%
\bibitem{Herzog:1995}%
  \BibitemOpen
  \bibfield{author}{%
  \bibinfo {author} {\bibfnamefont{T.~J.}\ \bibnamefont{Herzog}}, \bibinfo
  {author} {\bibfnamefont{P.~G.}\ \bibnamefont{Kwiat}}, \bibinfo {author}
  {\bibfnamefont{H.}~\bibnamefont{Weinfurter}},\ and\ \bibinfo {author}
  {\bibfnamefont{A.}~\bibnamefont{Zeilinger}},\ }%
  \bibfield{journal}{%
  \bibinfo {journal} {Phys. Rev. Lett.}\ }%
  \textbf{\bibinfo {volume} {75}},\ \bibinfo {pages} {3034} (\bibinfo {year}
  {1995})%
  \bibAnnoteFile{NoStop}{Herzog:1995}%
\bibitem{Kim:2000}%
  \BibitemOpen
  \bibfield{author}{%
  \bibinfo {author} {\bibfnamefont{Y.-H.}\ \bibnamefont{Kim}}, \bibinfo
  {author} {\bibfnamefont{R.}~\bibnamefont{Yu}}, \bibinfo {author}
  {\bibfnamefont{S.~P.}\ \bibnamefont{Kulik}}, \bibinfo {author}
  {\bibfnamefont{J.}~\bibnamefont{Shih}},\ and\ \bibinfo {author}
  {\bibfnamefont{M.~O.}\ \bibnamefont{Scully}},\ }%
  \bibfield{journal}{%
  \bibinfo {journal} {Phys. Rev. Lett.}\ }%
  \textbf{\bibinfo {volume} {84}},\ \bibinfo {pages} {1} (\bibinfo {year}
  {2000})%
  \bibAnnoteFile{NoStop}{Kim:2000}%
\bibitem{Englert:2000}%
  \BibitemOpen
  \bibfield{author}{%
  \bibinfo {author} {\bibfnamefont{B.-G.}\ \bibnamefont{Englert}}\ and\
  \bibinfo {author} {\bibfnamefont{J.~A.}\ \bibnamefont{Bergou}},\ }%
  \bibfield{journal}{%
  \bibinfo {journal} {Opt. Comm.}\ }%
  \textbf{\bibinfo {volume} {179}},\ \bibinfo {pages} {337} (\bibinfo {year}
  {2000})%
  \bibAnnoteFile{NoStop}{Englert:2000}%
\bibitem{Note1}%
  \BibitemOpen
  \bibinfo {note} {In the particular case $\protect \mathcal V = 1$, there is
  of course a vanishing probability to detect a quanton in the minima of the
  interference pattern.}%
  \bibAnnoteFile{Stop}{Note1}%
\end{thebibliography}%
\end{document}